\documentclass[aps,prb,reprint,showpacs,floatfix,citeautoscript,superscriptaddress,longbibliography]{revtex4-2}

\usepackage{graphicx}
\usepackage{dcolumn}
\usepackage{bm}
\usepackage{amsthm,amsmath,amssymb,mathrsfs}
\usepackage[colorlinks=true, linkcolor=blue, citecolor=blue, urlcolor=blue, linktoc=page, bookmarks=false]{hyperref}
\usepackage{color}
\usepackage{epstopdf}
\usepackage{subfigure}
\usepackage{units}
\usepackage{multirow}
\usepackage{upgreek}
\usepackage{booktabs}
\begin{document}

\renewcommand{\thefootnote}{\fnsymbol{footnote}}
\title{Stochastic \emph{p}-Bits Based on Spin-Orbit Torque Magnetic Tunnel Junctions}

\author{X. H. Li}
\affiliation{
	Beijing National Laboratory for Condensed Matter Physics, Institute of Physics, University of Chinese Academy of Sciences, Chinese Academy of Sciences, Beijing 100190, China
}%
\affiliation{
	Center of Materials Science and Optoelectronics Engineering, University of Chinese Academy of Sciences, Beijing 100049, China
}%

\author{M. K. Zhao}%
\affiliation{
	Beijing National Laboratory for Condensed Matter Physics, Institute of Physics, University of Chinese Academy of Sciences, Chinese Academy of Sciences, Beijing 100190, China
}%

\author{R. Zhang}%
\affiliation{
	Beijing National Laboratory for Condensed Matter Physics, Institute of Physics, University of Chinese Academy of Sciences, Chinese Academy of Sciences, Beijing 100190, China
}%

\author{C. H. Wan}%
\email{wancaihua@iphy.ac.cn}
\affiliation{
	Beijing National Laboratory for Condensed Matter Physics, Institute of Physics, University of Chinese Academy of Sciences, Chinese Academy of Sciences, Beijing 100190, China
}%

\affiliation{
	Songshan Lake Materials Laboratory, Dongguan, Guangdong 523808, China
}%

\author{Y. Z. Wang}%
\affiliation{
	Beijing National Laboratory for Condensed Matter Physics, Institute of Physics, University of Chinese Academy of Sciences, Chinese Academy of Sciences, Beijing 100190, China
}%

\author{X. M. Luo}%
\affiliation{
	Beijing National Laboratory for Condensed Matter Physics, Institute of Physics, University of Chinese Academy of Sciences, Chinese Academy of Sciences, Beijing 100190, China
}%

\author{S. Q. Liu}%
\affiliation{
	Beijing National Laboratory for Condensed Matter Physics, Institute of Physics, University of Chinese Academy of Sciences, Chinese Academy of Sciences, Beijing 100190, China
}%

\author{J. H. Xia}%
\affiliation{
	Beijing National Laboratory for Condensed Matter Physics, Institute of Physics, University of Chinese Academy of Sciences, Chinese Academy of Sciences, Beijing 100190, China
}%

\author{G. Q. Yu}%
\affiliation{
	Beijing National Laboratory for Condensed Matter Physics, Institute of Physics, University of Chinese Academy of Sciences, Chinese Academy of Sciences, Beijing 100190, China
}%

\affiliation{
	Songshan Lake Materials Laboratory, Dongguan, Guangdong 523808, China
}%

\author{X. F. Han}%
\email{xfhan@iphy.ac.cn}
\affiliation{
	Beijing National Laboratory for Condensed Matter Physics, Institute of Physics, University of Chinese Academy of Sciences, Chinese Academy of Sciences, Beijing 100190, China
}%

\affiliation{
	Center of Materials Science and Optoelectronics Engineering, University of Chinese Academy of Sciences, Beijing 100049, China
}%

\affiliation{
	Songshan Lake Materials Laboratory, Dongguan, Guangdong 523808, China
}%

\date{\today}

\begin{abstract}
Stochastic \emph{p}-Bit devices play a pivotal role in solving NP-hard problems, neural network computing, and hardware accelerators for algorithms such as the simulated annealing. In this work, we focus on Stochastic \emph{p}-Bits based on high-barrier magnetic tunnel junctions (HB-MTJs) with identical stack structure and cell geometry, but employing different spin-orbit torque (SOT) switching schemes. We conducted a comparative study of their switching probability as a function of pulse amplitude and width of the applied voltage. Through experimental and theoretical investigations, we have observed that the Y-type SOT-MTJs exhibit the gentlest dependence of the switching probability on the external voltage. This characteristic indicates superior tunability in randomness and enhanced robustness against external disturbances when Y-type SOT-MTJs are employed as stochastic \emph{p}-Bits. Furthermore, the random numbers generated by these Y-type SOT-MTJs, following XOR pretreatment, have successfully passed the National Institute of Standards and Technology (NIST) SP800-22 test. This comprehensive study demonstrates the high performance and immense potential of Y-type SOT-MTJs for the implementation of stochastic \emph{p}-Bits.
\end{abstract}
\maketitle
\section{Introduction}
With the culmination of Moore's Law and the advent of the big data era, traditional deterministic computing is facing challenges, particularly in terms of memory wall. Stochastic \emph{p}-bits have emerged as powerful tools for addressing Non-deterministic-Polynomial-hard (NP-hard)  problems, reversible reasoning, and neural network computing, and are poised to become the next generation of intelligent computing devices \cite{ref26,ref27,ref28,ref34,ref18,ref40,ref41,ref42,ref43,ref44,ref45,ref46,ref47}.

Among various technical approaches, spintronic stochastic \emph{p}-Bits based on thermal fluctuations offer advantages and show promise in terms of long endurance, low power consumption, compatibility, and scalability with advanced CMOS technologies. These devices can be seamlessly integrated into processors \cite{ref10,ref11,ref12,ref13,ref14,ref15,ref16,ref33}. A spintronic stochastic \emph{p}-Bit primarily consists of a magnetic tunnel junction (MTJ) in which the resistance randomly switches between a high resistance (antiparallel state) and a low resistance (parallel state) under specific critical conditions. These stochastic \emph{p}-Bits can be classified into three categories: low-barrier MTJs (LB-MTJs) \cite{ref17,ref19,ref33,ref37}, spin-transfer-torque-driven (STT) high-barrier MTJs (HB-MTJs) \cite{ref36,ref38,ref39}, and spin-orbit-torque-driven (SOT) HB-MTJs \cite{ref20,ref21,ref22,ref23,ref24,ref25}. While the randomness in all these devices arises from the influence of thermal fluctuations on the magnetization switching dynamics, there are significant differences in the energy barriers separating the parallel and antiparallel states of MTJs. LB-MTJs possess energy barriers comparable to \emph{$k_{\rm B}T$}, and their resistances can autonomously fluctuate without external stimulation, resulting in extremely low power consumption. These devices have been creatively employed as \emph{p}-bits in an integer factoring processor \cite{ref26,ref27,ref28}. In stark contrast, HB-MTJs feature much larger energy barriers than \emph{$k_{\rm B}T$}, and randomness can only be induced by thermal fluctuations when they are pre-driven via STT or SOT to specific critical states. Although they consume more energy, HB-MTJ stochastic \emph{p}-Bits can be systematically driven by clock signals, and the generated random numbers can be stored non-volatilely, enabling convenient data calling.

Despite occupying a larger cell size due to their three-terminal geometry, SOT-MTJs can, in principle, provide higher endurance, lower power consumption, and faster operation speed compared to their STT counterparts\cite{ref48}, making them well-suited for stochastic \emph{p}-Bit applications. Although various types of field-free SOT-MTJs have been experimentally demonstrated and optimized for memory applications \cite{ref29,ref30,ref31}, their performance in generating random numbers, tunability of randomness under external conditions, and susceptibility to applied stimuli have not been reported. This motivates our systematic study of stochastic \emph{p}-Bits based on various types of SOT-MTJs. Here, we fabricated Y-type, X-type, and XY-mixed type SOT-MTJs using identical stack and cell structures, and experimentally investigated their switching probability as a function of applied voltage and pulse duration. We have uncovered the underlying physics behind these diverse stochastic \emph{p}-Bit devices. Our results demonstrate that Y-type HB-SOT-MTJs outperform the other configurations as stochastic \emph{p}-Bits, offering greater tunability of randomness, lower power consumption, and tremendous potential for stochastic \emph{p}-Bit applications.

\section{Experiments}
In this letter, our MTJ stack structure is W(3)/$\rm{Co}_{20}\rm{Fe}_{60}\rm{B}_{20}$(1.6)/MgO(1.8)/$\rm{Co}_{20}\rm{Fe}_{60}\rm{B}_{20}$(1.8) /Co(1)/Ru(0.8)/Co(3)/IrMn(10)/Ru(4 nm), which was deposited by a magnetron sputtering system (ULVAC) with a base pressure of $1\times{10}^{-6} $ Pa. The numbers in brackets are thickness in nanometers. Due to an in-plane (IP) magnetic field of 180 Oe in the sputtering system, we can induce an easy axis during deposition. After a high temperature annealing process with a base vacuum of $1\times{10}^{-4} $ Pa, we measured the hysteresis loops of the stack by a vibrating sample magnetometer (VMS, EZ-9 from MicroSense) to check its in-plane easy and hard axes. The following is the detailed fabrication process of MTJs: (1) Use electron-beam lithography (EBL) to transfer the writing line pattern to films and the writing line is etched by ion beam etching (IBE) for the first time. (2) Use EBL to transfer the MTJ pattern to the writing line and IBE carves the junction. (3) $\rm{SiO}_2$ is grown to prevent side wall oxidation. (4) Ultraviolet lithography is used to pattern big electrode pads and Pt/Au films are deposited for external connections. Transport measurement is carried out on a three-dimensional probe platform with Keithley 2400, 2182A and Agilent 81104A. More details of the device fabrication and measurement have been reported elsewhere \cite{ref32}.

\section{Results and Discussions}

\begin{figure}[thb!]

\includegraphics[width=8.5cm]{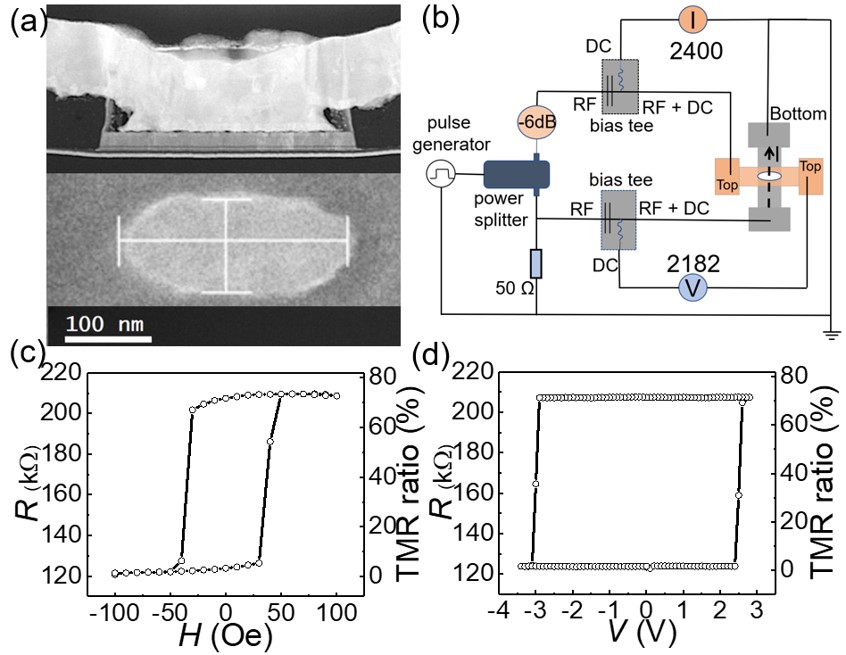}
\caption{\label{fig 1} \raggedright (a) High resolution high-angle angular dark field (HAADF) images of MTJ (top) and the SEM image of a Y-type MTJ (bottom). (b) Setup for the nanosecond pulse measurement. (c) Field-dependence of the resistance of SOT-MTJ. (d) Voltage-dependence of the resistance of the corresponding SOT-MTJ. }
\end{figure}

The structure of the MTJ device was confirmed by the high resolutional high-angle angular dark field (HAADF) image of the cross-section of a Y-type SOT-MTJ as shown in top of Fig.\ref{fig 1}(a). Its elementary mapping was also confirmed similar as the designed stack (not shown here). In the bottom of Fig.\ref{fig 1}(a), a SEM image of a Y-type SOT-MTJ after lift-off is shown. Its long and short axes are about 270 nm and 110 nm, respectively, which follows our pattern design. Measurement setup for SOT-MTJs is shown in Fig.\ref{fig 1}(b), which permits us to source as short pulses as  $\geqslant 10$ ns into the writing line. MTJ resistances were read using a 0.1 $\rm{\mu}$A current from a Keithley2400 current source and using a Keithley2182 nanovoltmeter to measure voltages cross MTJs. An Agilent 81104A functioned as a nanosecond pulse generator to source pulse currents into writing lines. In order to prevent STT disturbance and breakdown of MgO barriers caused by a high voltage pulse, the pulse was divided into two paths: one flowed through the tungsten writing channel to switch the free layer of a SOT-MTJ, and the other was attenuated by -6 dB (about half of the other in amplitude) and connected with the top electrode to minimize the dropped voltage across a MTJ. Fig.\ref{fig 1}(c) and (d) show the field- and voltage-dependences of the resistance of a Y-type SOT-MTJ. Their TMR ratio is both about 70\%, indicating nearly full magnetization switching. Sharp transitions between the parallel (P) and antiparallel (AP) states have been observed.

\begin{figure}[thb!]
\includegraphics[width=8.5cm]{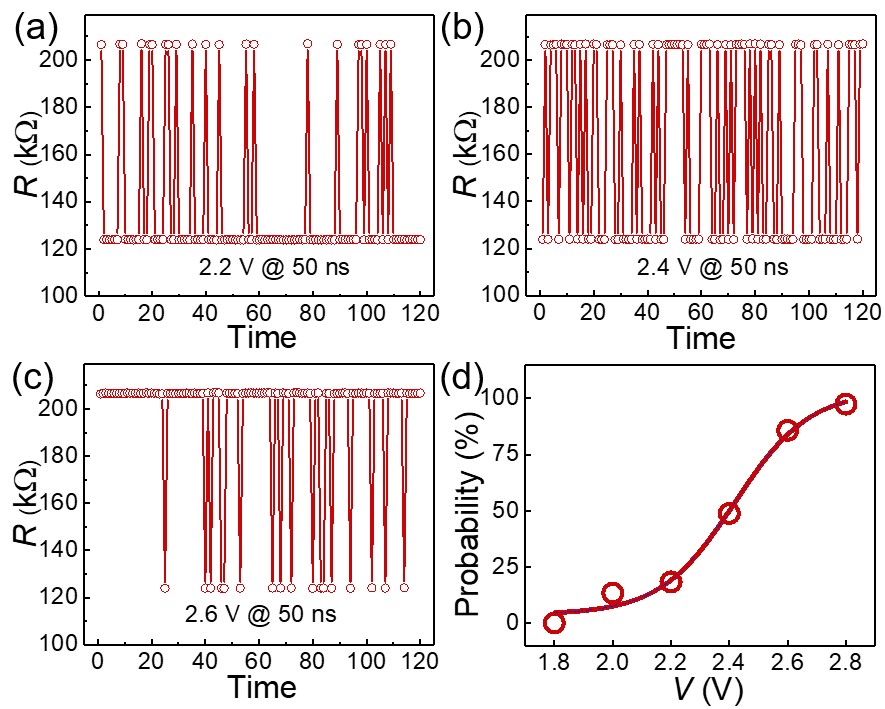}
\caption{\label{fig 2} \raggedright A stochastic \emph{p}-Bit device demonstration using the Y-type SOT-MTJ. Resistance of the Y-type SOT-MTJ at $V_{in}$=2.2 V (a) , 2.4 V (b) and 2.6 V (c) after 120 writing pulses. (d) Switching probability as a function of $V_{in}$ with a well-matched sigmoid curve.}  
\end{figure}

In order to characterize the \emph{p}-bit characteristics of the Y-type MTJs, we measured the switching probabilities (\emph{P}) under different voltages as shown in Fig.\ref{fig 2}(a-c). It shows the detailed output resistance after the voltage pulses of the same 50 ns pulse width but difference amplitudes 2.2 V (a), 2.4 V (b) and 2.6 V (c), respectively. Their corresponding switching probabilities are 20\%, 50\% and 80\% respectively. A sigmoid function is used to fit the voltage-dependence of the switching probability data and they excellently match as displayed in Fig.\ref{fig 2}(d). Over 15 Kb random numbers generated by the Y-type SOT-MTJ, after an XOR operation, have been checked out by the NIST SP800-22 test to confirm randomness of the random numbers and all items tested as shown in Table 1 have passed the examination (\emph{P}-value $>$ 0.01 and Proportion $>$ 0.97). These results manifest that (1) our Y-type MTJs exhibit excellent random switching features and (2) their switching probability can be conveniently tuned by applying voltages. Moreover, the voltage interval $\Delta V_c$ is about 0.4 V as the switching probability changes from 20\% to 80\%. Here, we define a ratio \emph{$Q$}=\emph{$\Delta V_c/V_c$}(\emph{P}=50\%) as the quantitative merit value of a SOT-MTJ because the larger \emph{Q} indicates a wider voltage tunability, a gentler susceptibility to the external stimulus and a lower energy consumption. And the ratio \emph{Q} reaches 17\% for our Y-type SOT-MTJs, which can provide a wide voltage margin of circuitry design for voltage-controllable stochastic \emph{p}-Bit devices.

\newcommand{\tabincell}[2]{\begin{tabular}{@{}#1@{}}#2\end{tabular}}  

\begin{table}[!htbp]
\caption{ Randomness test (NIST 800-22 test)}
\begin{tabular}{|c|c|c|c|c|c|c|}
\hline
 No.& Test&\multicolumn{2}{c|}{P-value}&\multicolumn{2}{c|}{Proportion}&Success/Fail\\
\hline
1&Frequency&\multicolumn{2}{c|}{0.34} &\multicolumn{2}{c|}{1}&Success\\
\hline
2&Block Frequency&\multicolumn{2}{c|}{0.53}&\multicolumn{2}{c|}{1}&Success\\
\hline
3&Runs&\multicolumn{2}{c|}{0.31}&\multicolumn{2}{c|}{1}&Success\\
\hline
4&\tabincell{c}{Longest \\ Run of Ones} &\multicolumn{2}{c|}{0.58}&\multicolumn{2}{c|}{1}&Success\\
\hline
5&FFT&\multicolumn{2}{c|}{0.10}&\multicolumn{2}{c|}{1}&Success\\
\hline
6&\tabincell{c}{Approximate \\ Entropy}  &\multicolumn{2}{c|}{0.09}&\multicolumn{2}{c|}{1}&Success\\
\hline
7&Cumulative Sums&0.61&0.48&\multicolumn{2}{c|}{1}&Success\\
\hline
8&Serial&0.48&0.49&\multicolumn{2}{c|}{1}&Success\\
\hline
9&\tabincell{c}{Overlapping \\ Template} &\multicolumn{2}{c|}{0.75}&\multicolumn{2}{c|}{1}&Success\\
\hline
10&\tabincell{c}{Non-overlapping \\ Template Matching}& \multicolumn{2}{c|}{ } &\multicolumn{2}{c|}{0.973}&Success\\
\hline
11&Rank&\multicolumn{2}{c|}{0.48}&\multicolumn{2}{c|}{1}&Success\\
\hline
12&Linear Complexity&\multicolumn{2}{c|}{0.24}&\multicolumn{2}{c|}{1}& Success\\
\hline
\end{tabular}
\end{table}

\begin{figure*}[thb!]
\centering
\includegraphics[width=16cm]{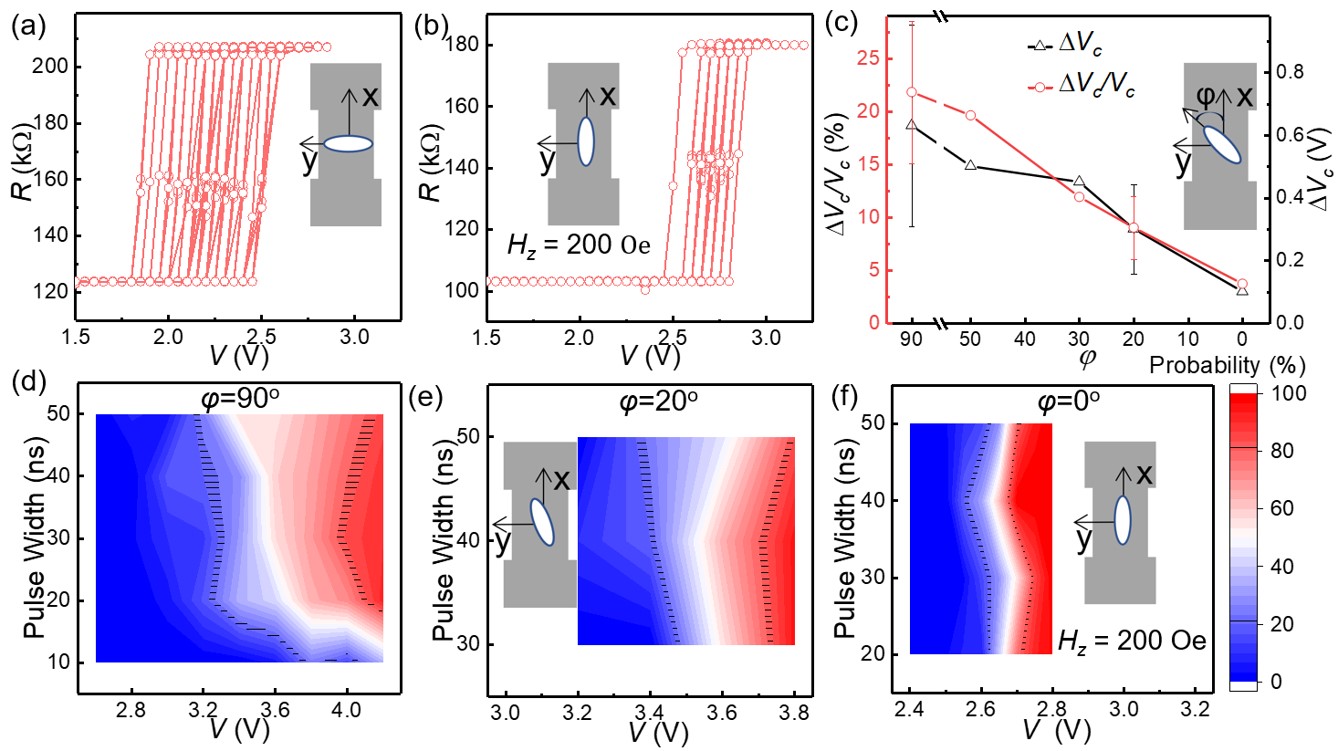}
\caption{\label{fig 3} \raggedright Comparison of the switching probability performance of the Y-type, XY-mixed type and X-type SOT-MTJs. (a) and (b) The switching performance of the Y-type and X-type SOT-MTJs, respectively, measured for 50 times at the 50 ns pulse width. (c) The angle (\emph{$\varphi$})-dependence of the ratio ($\Delta V_c/V_{c,\emph{P}=50\%}$). \emph{$\varphi$} =$90^{\circ}$ ($0^{\circ}$) corresponds to the Y-type (X-type) SOT mode and the other values denote the mixed XY mode. \emph{$\Delta V_c$} $\equiv$ \emph{$V_c$}(\emph{P}=80\%)-\emph{$V_c$}(\emph{P}=20\%) is captured as the switching probability \emph{P} changes from 20\% to 80\%. The ratio is obtained by dividing \emph{$\Delta V_c$} with the critical voltage \emph{$V_c$} at \emph{P}=50\%. (d, e and f) The phase diagrams of the switching probability of the Y-type, XY-mixed type (\emph{$\varphi$} =$20^{\circ}$) and X-type SOT-MTJs as functions of different pulse widths and voltage amplitudes. The 20\% and 80\% switching probabilities are highlightened by the dashed lines accordingly. Note that the interval of the x axis (the \emph{V} axis) of (e) and (f) is half of that of (d). }
\end{figure*}

As mentioned above, when designing peripheral circuits for a random number generator, the defined merit factor \emph{Q} is critical in the sense of power consumption, precision requirements of the circuits and robustness to circuitry disturbances. Especially, for a stochastic \emph{p}-Bit device with a specified 1/0 output probability, a too narrow switching interval \emph{$\Delta V_c$} will raise a redundant challenge for a voltage-stabilizing circuit. To achieve a higher \emph{Q} factor, therefore, we systematically studied the switching probability of various types of SOT-MTJs with different patterns but the same stack and unit cell structures. Fig.\ref{fig 3}(a) and (b) show the raw switching curves of the Y-type and X-type SOT-MTJs at a 50 ns pulse width for 50 repetitions. A device is first initialized at its parallel state. Then, the dependence of the MTJ resistance as a function of the amplitude of a voltage pulse is traced for 50 times. It can be intuitively identified that the switching interval of the Y-type SOT-MTJ is significantly larger than that of the X-type SOT-MTJ. For characterizing switching phase diagrams in more details, we also measured the switching probability \emph{P} under different pulse widths. Fig.\ref{fig 3}(d) shows the phase diagram of the Y-type SOT-MTJ from 10 ns to 50 ns.  Several SOT-MTJs of the Y-type scheme all show gentler voltage dependence of the switching probability. At the first glance, the critical voltage \emph{$V_c$} at \emph{P}=50\% increases sharply as the pulse width down to 10 ns from 20 ns, which is a typical feature of the damping-dominated switching for the Y-type SOT scheme and the classic STT mode as well. Above 20 ns, the critical switching voltage tends stable. In the following, we mainly compare the switching properties (especially the \emph{Q} factor) of different types of SOT-MTJs in this region. 

We can see in Fig.\ref{fig 3}(d) that the switching interval \emph{$\Delta V_c$} is almost unchanged from 20 ns to 50 ns as well as \emph{$V_c$}(\emph{P}=50\%). Fig.\ref{fig 3}(e) and (f) show, correspondingly, the switching phase diagrams of the XY-mixed type (\emph{$\varphi$} = $20^{\circ}$) and X-type SOT-MTJs. The definition of the angle \emph{$\varphi$} can be found in the inset of Fig.\ref{fig 3}(c). It is defined in the XOY plane between the long axis of MTJ ellipses with the direction of current density. It can be straightforwardly found in Fig.\ref{fig 3} (d)-(f) that the switching interval \emph{$\Delta V_c$} increases rapidly with the increase in \emph{$\varphi$}. This trend is marked by two dashed equal probability lines corresponding to \emph{P}=20\% and 80\% in the figures. In order to describe more clearly and quantitatively the \emph{Q} factor of different types of SOT-MTJs, we summarize the \emph{$\varphi$}-dependence of \emph{Q} in Fig.\ref{fig 3}(c). Here \emph{$\varphi$}=$0^{\circ}$, $20^{\circ}$, $30^{\circ}$, $50^{\circ}$ and $90^{\circ}$. Remarkably, the \emph{Q} factor as well as \emph{$\Delta V_c$} both systematically declines with decreasing \emph{$\varphi$}. It is worth stressing that the data are acquired from various types of SOT-MTJs but with the same stack and unit cell structures. Therefore, this set of data can be instructive to find suitable modes for different application scenarios. Besides, different stochastic performances should be more probably attributed to the various SOT modes since their stack and cell structure is the same.  

\begin{figure*}[thb!]
\centering
\includegraphics[width=16cm]{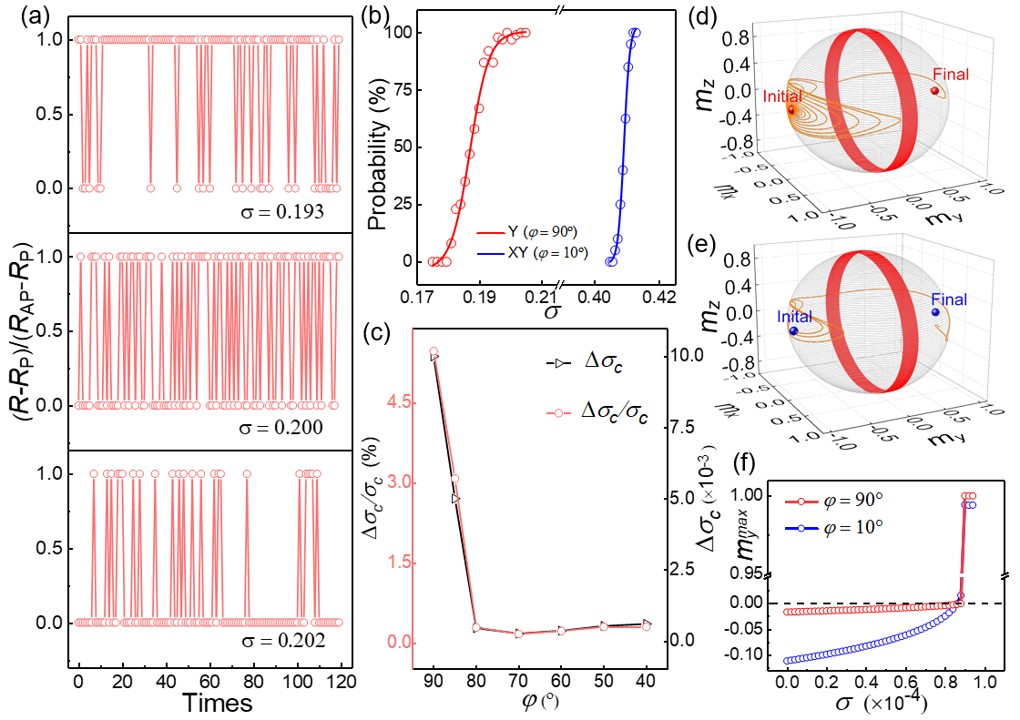}
\caption{\label{fig 4} \raggedright Macrospin simulation results of a stochastic \emph{p}-Bit device based on a simulated Y-type SOT-MTJ. (a) The simulated probability results of the Y-type SOT-MTJ working at different probabilities by setting different spin-orbit torques.  (b) The switching probability as a function of the applied spin torque \emph{$\sigma$} for the Y-type (\emph{$\varphi$}=$90^{\circ}$) and XY-mixed type (\emph{$\varphi$}=$10^{\circ}$) of SOT-modes. Sigmoid functions are discovered well matched with the calculated voltage-dependence of the switching probability. Apparently, the calculated Q and \emph{$\Delta \sigma_c$} is significantly higher in the Y-type SOT-MTJs (\emph{$\varphi$}=$90^{\circ}$). (c) The \emph{$\varphi$}-dependence of the calculated \emph{Q} factor, which qualitatively reproduces the observed trend in Fig.\ref{fig 3}(c). The magnetization switching dynamics driven by SOT of the Y-type with (d) \emph{$\varphi$}=$90^{\circ}$ and (e) the XY-mixed type (\emph{$\varphi$}=$10^{\circ}$). The red areas show ridges in the energy profile to determine the switching or not. The red stripes correspond to the regions with -0.1$<$\emph{$m_y$}$<$0.1. (f) The spin torque (\emph{$\sigma$})-dependence of \emph{$m_y^{max}$} for the Y-type and XY-mixed type dynamics.}
\end{figure*}

In order to understand our experimental results in various types of SOT-MTJs, we utilized a macrospin model based on the generalized LLGS equation to simulate spin dynamics and switching probability. The classic Landau-Lifshitz-Gilbert equation is enriched by the Slonczewski-like spin-transfer torque term or the so-called damping-like torque term as shown in Equation (1). A random field with its amplitude proportional to \emph{$k_{\rm{B}}T$} and its direction fully random is imposed additionally on the system as a result of the thermal agitation effect which is deemed as the nature source of randomness for MTJ-based TRNGs. The model is further numerically solved.

\begin{equation}
\frac{\partial\vec{m}}{\partial t}=-\gamma\vec{m}\times\left({\vec{H}}_{\rm{eff}}+{\vec{H}}_{\rm{T}}\right)+\alpha\vec{m}\times\frac{\partial\vec{m}}{\partial t}+\vec{m}\times\vec{\sigma}\times\vec{m}                                \tag{1}
\end{equation}
where $\vec{m}$ is the unit vector of magnetization, \emph{$\gamma$} is the gyromagnetic ratio, \emph{$\alpha$} is Gilbert damping constant, $\vec{\sigma}$ is the vector of an applied spin torque, $\mathbf{H}_{eff}$ is an effective field including an in-plane anisotropy field introduced by the induced shape anisotropy and an out-of-plane anisotropic field due to demagnetization in the present case and $\mathbf{H}_T$ is the random thermal field. Worth noting, we only consider the Y-type and XY-mixed type SOT modes in the simulation because these two modes support field-free operations. As mentioned before, $\mathbf{H}_T$ at a fixed temperature is assumed constant in its amplitude but fully random in its direction. Note that the duration of $\mathbf{H}_T$ is set the same as the time interval for the spin dynamic evolution process. We have also simulated the influence of the field-like torque in the supporting material and the result indicates that the effect of the field-like torque is not the main factor affecting the switching interval compared with the effect of various SOT schemes.

For the Y-type SOT-MTJ, when $\mathbf{H}_T$ is switched off, the spin dynamics solved by Equation (1) is trivial and shown in inset of Fig.\ref{fig 4}(d), no difference with others \cite{ref31}. The magnetization switches from -y to +y after imposing a large enough spin torque along the +y direction on it. After introducing $\mathbf{H}_T$, we obtain the 20\%, 50\% and 80\% switching probabilities under elevated torque amplitudes as shown in Fig.\ref{fig 4}(a), which is consistent with our experimental results in Fig.\ref{fig 2}(a). Fig.\ref{fig 4}(b) summarizes the switching probability at different applied torques in the case of Y-type (\emph{$\varphi$}=$90^{\circ}$) and XY-mixed type (\emph{$\varphi$}=$10^{\circ}$) modes and also their fitting results by the sigmoid functions. The Y-type SOT-MTJ in the simulation has the higher interval \emph{$\Delta \sigma_c$} and the lower critical torque \emph{$\sigma_c$} than the XY-mixed type as clearly manifested in Fig.\ref{fig 4}(b). Finally, we calculate the merit factor \emph{Q} (defined as \emph{$\Delta \sigma_c /\sigma_c$}  in simulation) at different \emph{$\varphi$} as shown in Fig.\ref{fig 4}(c). As the \emph{$\varphi$} decreases from the Y-type (\emph{$\varphi$}=$90^{\circ}$) to the XY-mixed type (\emph{$\varphi$} $\in 80^{\circ}$$\sim$$40^{\circ}$), the switching interval \emph{$\Delta \sigma_c$} and the merit factor \emph{Q} both decrease remarkably, reproducing qualitatively our experimental results in Fig.\ref{fig 3}(c).

In Equation (1), because of much smaller \emph{$k_{\rm{B}}T$} than the energy barrier for the HB-MTJs, the random field $\mathbf{H}_T$ can be deemed as a perturbation to $\mathbf{H}_{eff}$ and the mildness of the voltage-dependence of switching probability for the Y-type SOT-MTJs should be evidenced by the spin dynamics with $\mathbf{H}_T$=0. To account for the gentler voltage-dependence of the switching probability in the Y-type SOT-MTJ than the XY-mixed type, we thus traced a parameter \emph{$m_y^{max}$}, the maximized magnetization component in the y axis (the easy axis) during the dynamic process activated by a specified spin torque \emph{$\sigma$} . Since for a HB-MTJ stochastic \emph{p}-Bit device with energy barriers $\gg$ \emph{$k_{\rm{B}}T$}, only the events with $m_y^{max}$ approaching zero (the ridge in the energy profile to determine the switching or not as marked by the red area in Fig.\ref{fig 4}(d) and (e)) can allow thermal fluctuations in the order of \emph{$\sim k_{\rm{B}}T$} involved and play a critical role in the switching process from \emph{$m_y$} = -1 to +1. This is the reason why we interest in the \emph{$m_y^{max}(\sigma)$} relation as shown in Fig.\ref{fig 4}(f). This figure clearly shows the Y-type SOT scheme has a much lower slope \emph{$S$}$\equiv$ $\rm{d}$$m_{y}^{max}/\rm{d}\sigma $ around \emph{$m_y^{max}$}=0 between \emph{$\sigma$} = \emph{$\sigma_c$} and \emph{$\sigma$} = \emph{$\sigma_c$}-8.4$\times{10}^{-5}$. A smaller \emph{S} enables the magnetization of  the free layer to dwell for a longer time near the regions of $m_y$=0 where the magnetization is more vulnerable to thermal fluctuations. Thus the smaller \emph{S} should be desired to reduce the voltage sensitivity of the switching probability and flatten the \emph{P}-\emph{V} sigmoid dependence. Here, for straightforward comparison, we have already offset their \emph{$\sigma_c$} to the same position by subtracting a constant \emph{$\sigma_c$} in Fig.\ref{fig 4}(f), which does not affect the slope value of \emph{S}. Given the same magnetic parameters, \emph{$S$} $\approx$ 200 and 1200 for the Y-type and XY-mixed type schemes, respectively. This huge difference in \emph{S} persuades us in physics why the Y-type scheme shows the smoother voltage-tunability on switching probability.

The experimental \emph{Q} factor $\sim$17\% is about 3 times as large as the calculated one because in reality such magnetic parameters as the anisotropic fields and damping constant may have some spatial distributions among different places of an MTJ unit and this uncertainty can surely broaden the \emph{$\Delta V_c$} and enlarge the \emph{Q}. But even such a simple toy model can grasp the huge contrast in \emph{Q} (from $\sim$5\% to below 1\%, higher than 5 times) as changing the SOT-MTJ type from the Y-type mode to the XY-mixed type. These experimental and simulation results both clearly manifest the importance of selecting suitable SOT modes for targeting various aims of SOT-devices. While a small interval \emph{$\Delta V_c$}  may be benefit for memory applications, the Y-type SOT-MTJs discovered with a higher \emph{Q} factor here is no doubt more favorable for the stochastic \emph{p}-Bit uses.

\section{Conclusion}
Our experiment shows that the switching probability can be accurately adjusted by the switching voltage, which is a critical factor for stochastic \emph{p}-Bit applications. Then, we systematically studied the switching interval \emph{$\Delta$}\emph{$V_c$} of the Y-type and XY-mixed type and X-type SOT switching modes in details. The switching features of various SOT modes can be grasped by a simple macrospin model taking a thermal field into account. The Y-type mode is found both in experiments and simulations to have the largest switching interval and low critical current density and thus the highest \emph{Q} factor, which eases circuitry design correspondingly and favorably supports its stochastic \emph{p}-Bit applications.

\section{Acknowledgements}
This work was supported by the National Key Research and Development Program of China (MOST) (Grant No. 2021YFB3601302 and 2017YFA0206200), the National Natural Science Foundation of China (NSFC) (Grant Nos. 51831012, 51620105004, and 11974398), the Strategic Priority Research Program (B) of Chinese Academy of Sciences (CAS) (Grant Nos. XDB33000000), and Beijing Natural Science Foundation (Grant No. Z201100004220006). C. H. Wan appreciates financial support from the Youth Innovation Promotion Association, CAS (Grant No. 2020008).  X.H. Li and M.K. Zhao contributed equally to this work. 

%

\end{document}